%% file: leff_mh.tex
\documentclass[5p,twocolumn,sort&compress]{elsarticle}

\journal{Physics Letters B}

\usepackage[utf8]{inputenc}
\usepackage{graphicx}
\usepackage{dcolumn}
\usepackage{bm}
\usepackage[colorlinks=false]{hyperref}
\usepackage{amsmath}
\usepackage{amsfonts}
\usepackage{amssymb}
\usepackage{wasysym}
\usepackage{MnSymbol}
\usepackage{ulem}
\renewcommand \xout{\bgroup \markoverwith{\kern-.15em{\scriptsize{/}}\kern-.15em}\ULon}
\newcommand \yout{\bgroup \markoverwith{\kern-.15em{\scriptsize{\textbackslash}}\kern-.15em}\ULon}
\newcommand \leff{$\mathcal{L}_{\mathrm{eff}}$}
\newcommand \qy{$\mathcal{Q}_{\mathrm{y}}$}
\newcommand \kevnr{$\mathrm{keV_{nr}}$}
\newcommand \kevee{$\mathrm{keV_{ee}}$}
\begin{document}

\title{Nuclear recoil scintillation and ionisation yields\\ in liquid xenon from {ZEPLIN--III} data}

\input author_elsa.tex

\begin{abstract}
Scintillation and ionisation yields for nuclear recoils in liquid xenon above $10\,$\kevnr (nuclear recoil energy) are deduced from data acquired using broadband Am-Be neutron sources.
The nuclear recoil data from several exposures to two sources were compared to detailed simulations.
Energy-dependent scintillation and ionisation yields giving acceptable fits to the data were derived.
Efficiency and resolution effects are treated using a light collection Monte Carlo, measured photomultiplier response profiles and hardware trigger studies.
A gradual fall in scintillation yield below $\sim40\,$\kevnr\ is found, together with a rising ionisation yield; both are in agreement with the latest independent measurements.
The analysis method is applied to the most recent {ZEPLIN--III} data, acquired with a significantly upgraded detector and a precision-calibrated Am-Be source, as well as to the earlier data from the first run in 2008.
A new method for deriving the recoil scintillation yield, which includes sub-threshold S1 events, is also presented which confirms the main analysis.
\end{abstract}
\maketitle

\renewcommand{\thefootnote}{\fnsymbol{footnote}}

\section{Introduction}
{ZEPLIN--III} \cite{sumner:2001um, Araujo:2006cr, Akimov:2007dm} is a dark matter search instrument for the direct detection of weakly interactive massive particles (WIMPs) via their elastic scattering from xenon target nuclei.
It records nuclear recoil events via two response channels: scintillation and ionisation.
A single array of photomultipliers (PMTs) records two light signals: one prompt, due to scintillation in the liquid (S1); and the other delayed, due to the electroluminescence of ionisation charge drifted into the gas region (S2). 
To use the two signals as energy estimators, and so infer the spectrum of a WIMP-nucleon scattering population, energy-dependent yields of scintillation and ionisation must be established. 
We adopt the conventional definitions: \leff($E$) is the scintillation yield for nuclear recoils of energy $E$ (commonly denoted as true nuclear recoil energy in \kevnr) relative to that of electron recoils of $122\,\mathrm{keV}$ photoabsorption at zero electric field; \qy($E$) is the yield of ionisation charge leaving the interaction site, in electrons per unit energy (independently of any other efficiency).\\

For a nuclear recoil with energy deposit $E$ in \kevnr\ the reconstructed energy $\langle\mathrm{S1}\rangle$ in \kevee\ is related to \leff\ by:
\begin{equation}\label{leffeq}
\langle\mathrm{S1}\rangle =  \frac{S_{\mathrm{nr}}}{S_{\mathrm{ee}}}\mathcal{L}_{\mathrm{eff}}\left(E\right)E,
\end{equation} 
where ${S_\mathrm{nr}}$ is the electric field suppression factor for nuclear recoils of energy ${E}$, and ${S_\mathrm{ee}}$ is the corresponding factor for electron recoils, both defined as unity at zero applied electric field.
Similarly, using the reconstructed energy of the ionisation signal $\langle\mathrm{S2}\rangle$ (as calibrated using electron recoils of $\mathrm{122\,keV}$), \qy is defined by:
\begin{equation}\label{qyeq}
\frac{\langle\mathrm{S2}\rangle}{W\ q_0/q(\mathbf{|E|)}} = \mathcal{Q}_{\mathrm{y}} \left( E \right) E,
\end{equation}
where $W$ is the energy required to produce an electron-ion pair in liquid xenon at infinite electric field ($W$-value) and the ratio $q(|\mathbf{E}|)/q_0$ represents the relative fraction of charge collected at a finite field.
Models and parameterisations of ${q(\mathbf{|E|})}$ are taken from Ref.\,\cite{Aprile:1991bg, Jaffe:1913gs}.\\

The present work measures both yield parameters by fitting simulations to observed scintillation and ionisation spectra for neutron calibration with an $\mathrm{^{241}Am}$-Be ($\alpha$,n) source. 
This spectral-matching approach has previously been applied by others to argon \cite{Benetti:2008kd} and to xenon \cite{Sorensen:2009ec, Sorensen:2010ex} and was also used by {ZEPLIN--III} in analysing its first science run \cite{Lebedenko:2009ce}.
Alternatively, measurements with mono-energetic neutrons can derive event-by-event recoil energies from the kinematics of the scattering angles.
Such beam measurements rely on small prototype chambers rather than larger WIMP detectors operating underground.
Several of these neutron-tagging measurements exhibiting a gradual fall below $40\,$\kevnr\ have been reported recently \cite{Plante:2011hw, Manzur:2010bf}.\\

We proceed by describing relevant details of event acquisition and selection from neutron calibration; the Monte Carlo simulation of calibration data; estimates of detector resolution and efficiency and their application to simulated events; and the process of spectral fitting to obtain \leff\ and \qy, including the resultant confidence intervals. 
We include spectral yield measurements from the first \cite{Lebedenko:2009ce} and second science run configurations of {ZEPLIN--III}, having re-analysed the earlier data with improved software.
The results are further consolidated by using an independent new method to analyse events below the scintillation threshold.
\leff\ is therein derived by attributing the average scintillation signals to the nuclear recoil energies reconstructed using the ionisation yield \qy.
The combined effects of the revised efficiency and light yield on WIMP-nucleon scattering limits are presented.

\section{Neutron data}
\begin{figure}[tb]
\includegraphics[width=\linewidth]{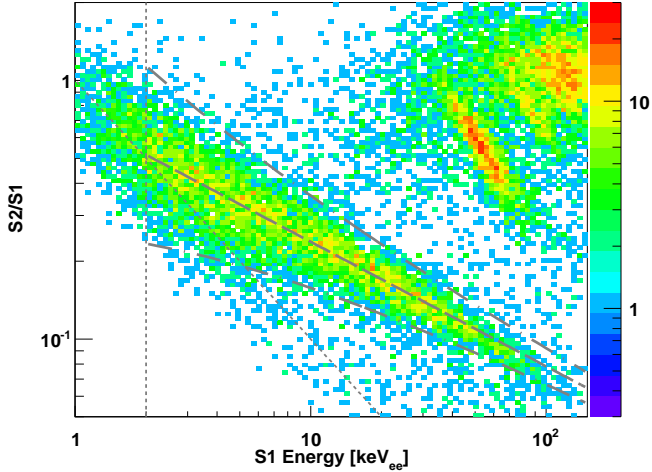}
\caption{\label{fig:ambe_loglog} The discrimination parameter $\mathrm{(S2/S1)}$ for the nuclear recoil response to an Am-Be neutron source. 
The colour scale indicates the counts for a $4.9\,\mathrm{h}$ exposure in the FSR configuration. 
The analysis thresholds for scintillation (S1) and ionisation (S2) are indicated by dotted lines. 
The centroid of the population and the $\pm$\,2\,$\sigma$ contour used for this analysis are indicated by the dashed and dash-dotted lines.
The $40$ and $80\,$\kevee\ inelastic populations can also be seen.}
\end{figure}
The first science run (FSR) configuration and data sets of {ZEPLIN--III} are described in detail in Ref.\,\cite{Lebedenko:2009ce}. 
The main neutron calibration data set was acquired in May 2008.
For the 2010-11 second science run (SSR) the PMTs were replaced by new units, with a 40-fold reduction in radioactivity \cite{Araujo:2011to} and an anti-coincidence (veto) detector was installed around the experiment \cite{Akimov:2010uv, Ghag:2011tx}.
Other changes include the use of a calibrated neutron source positioned centrally above the active volume and a $\sim10\%$ lower drift field in the liquid of $3.4\,\mathrm{kV\,cm}^{-1}$. 
The trigger for event acquisition was obtained from a shaped sum of PMT signals which, at low energies, was derived from the S2 signal.\\

\begin{figure}[tb]
$\begin{array}{c}
	\includegraphics[width=\linewidth]{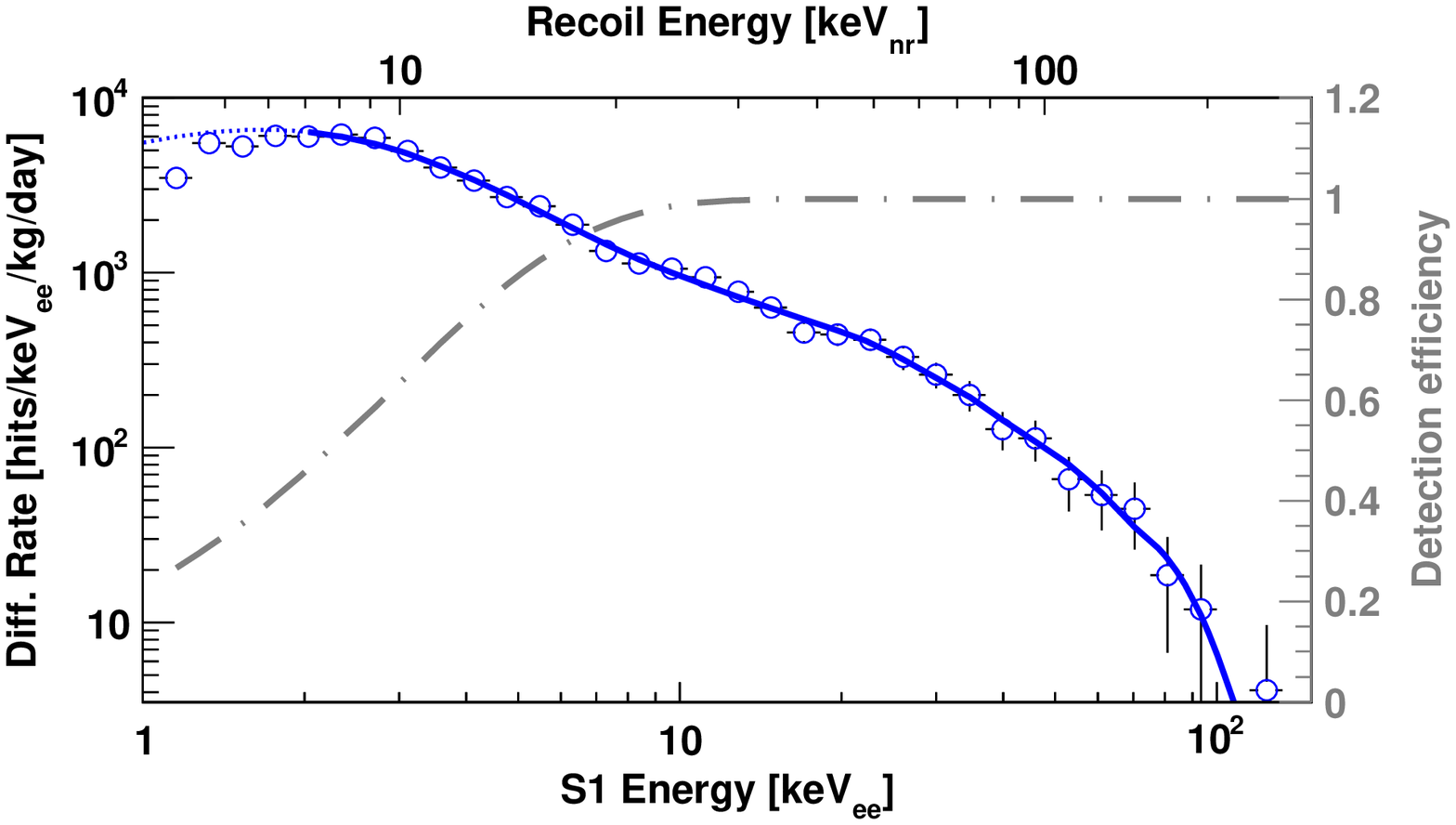}\label{fig:FSR_data_MC_leff} \\
	\includegraphics[width=\linewidth]{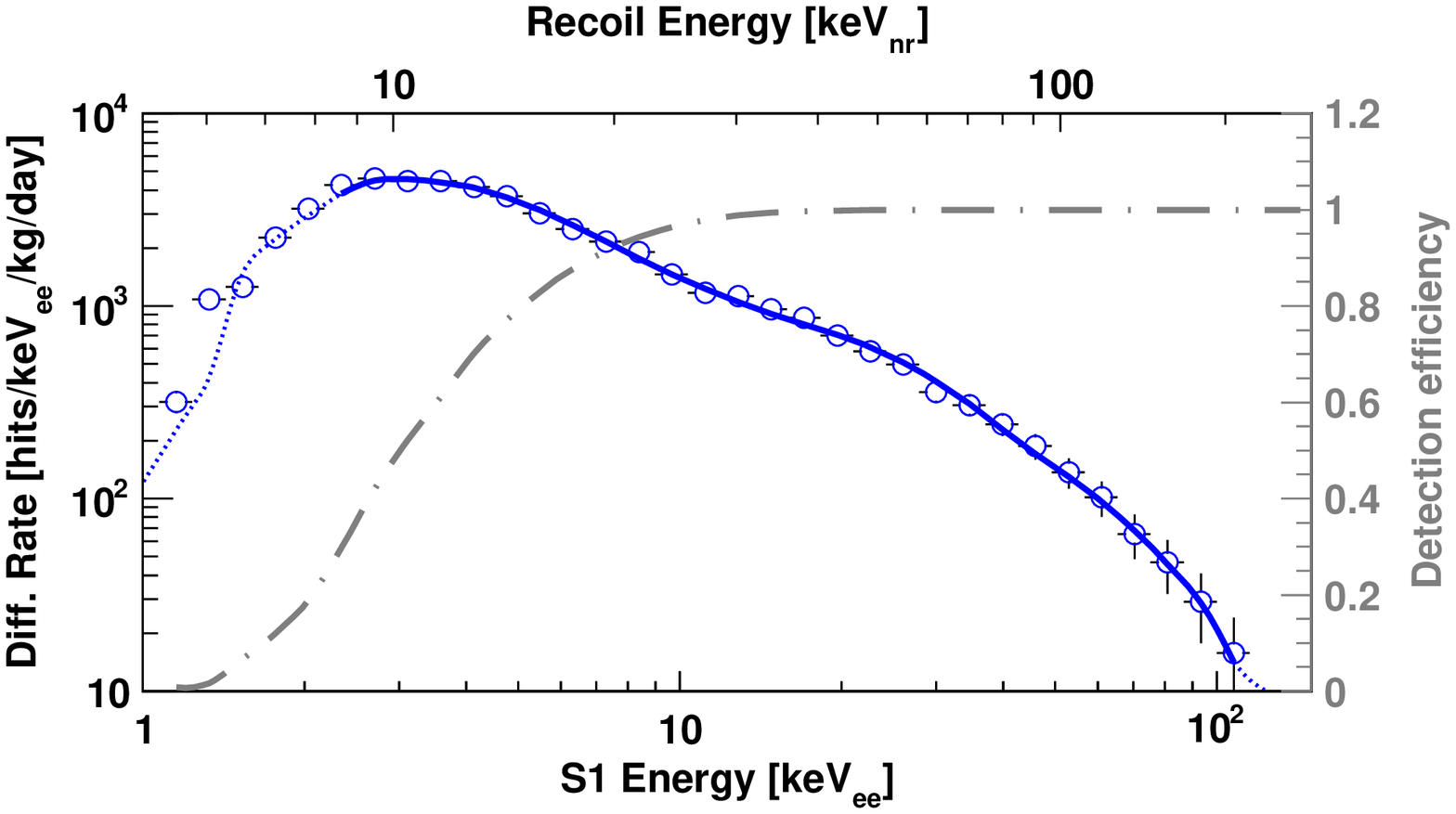}\label{fig:SSR_data_MC_leff}
\end{array}$
\caption{\label{fig:data_MC_leff} The differential spectrum of the primary scintillation of single scatter nuclear recoils from the 2008 (top) and 2010 (bottom) Am-Be data. 
Also shown are the results from a detailed Monte Carlo simulation (solid lines) considering an energy dependence of \leff\ as in Fig.\,\ref{fig:l_eff} and the corresponding nuclear recoil energy scale. 
S1 energies below $2.0\,$\kevee (FSR) and $2.5\,$\kevee (SSR) are not used to obtain the fit and are shown as extrapolations only (dotted). 
Superimposed is the detection efficiency (dash-dotted).}
\end{figure}

In the SSR, the Am-Be source was inserted through the experiment shielding via a dedicated delivery pipe, coming to rest above the cryostat vessel just $\sim5\,\mathrm{cm}$ offset from the PMT array centre.
The software acceptance threshold for S1 pulses requires a 3-fold coincidence amongst the 31 PMTs of at least one photoelectron in each channel.
In this analysis, only events representing single elastic scatters of neutrons are considered.
The event selection is achieved using only basic waveform and pulse quality cuts.
Interactions of photons from the source, inelastic scattering and radiative capture on xenon and other materials are rejected by event selection in $\mathrm{S2/S1}$, as shown in Fig.\,\ref{fig:ambe_loglog}.
Further selection criteria, e.g. to remove MSSI events (see Ref.\,\cite{Lebedenko:2009ce} for details), have been found to have no significant effect on the results of this analysis.
The SSR neutron data were acquired with a calibrated $\mathrm{20\,MBq}$ Am-Be source emitting $\mathrm{1321\pm14\,n/s}$\footnote{Calibration by National Physical Laboratory, Teddington, Middlesex, UK, May 2009} for a total of $\mathrm{10.2\,h}$ on three different days between June and August 2010.
A detailed analysis of the effect of the operational parameters showed no significant difference between the various datasets.
The primary scintillation spectrum of single elastic scatters within a $\mathrm{3.5\,kg}$ fiducial volume in the centre of the detector is shown for both FSR and SSR in Fig.\,\ref{fig:data_MC_leff}.\\

The SSR source was used to calibrate that employed in the first run using a low-background {HPGe} detector, resulting in $\mathrm{5512\pm358\,n/s}$.
In the FSR calibration (5-hour exposure) the neutron source was only marginally inserted into the shielding to avoid too high an event rate and irradiated the detector from one side, which changes the expected recoil spectrum slightly.
%
\section{Neutron scattering simulation}
Geant4 \cite{Allison:2006cd, Agostinelli:2003vd} Monte Carlo simulations of the entire {ZEPLIN--III} experiment have been implemented to predict the detector response to neutrons \cite{Araujo:2006cr}.
For the simulations in this work, Geant4.9.3. and updated Xe(n,n) scattering cross sections from the ENDF/B-VII database\,\cite{Chadwick:2006bm} were used.
A comprehensive and accurate setup of the experiment was implemented, taking into account the detector configuration and source position for the two data sets.
The simulations include the liquid and gaseous xenon, the PMT array, electrode grids, copper vessels, the source delivery mechanism, the plastic and lead shielding and, for the SSR configuration, also the veto detector.
The simulated energy spectrum for neutrons from the Am-Be source extends up to $\mathrm{11\,MeV}$ with a mean of $\mathrm{4.1\,MeV}$\,\cite{ISO8529-1:2001}.
The low-energy threshold was $\mathrm{100\,keV}$; no significant spectral uncertainty is to be expected above this energy for the low activity source considered here.
That corresponds to a maximum recoil energy of $\mathrm{3\,keV_{nr}}$ in xenon, which is significantly lower than the analysis threshold, and so this uncertainty has limited impact on the results.
Previous studies showed the Monte Carlo results for low energy nuclear recoils to be relatively insensitive to variations in the neutron source spectrum, the precise source location and the effect of intervening and surrounding materials\,\cite{Lebedenko:2009ce}.
The Monte Carlo event selection includes nuclear recoils from elastic and inelastic scattering (when prompt $\gamma$-rays may escape from the xenon undetected); 
recoils from radiative capture (when $\gamma$-rays can be delayed significantly) affect the spectrum only below 1\,$\mathrm{keV_{nr}}$ and were not considered.

\section{Detector response}
The overall detection efficiency is dependent on the quantum efficiency of the photomultipliers, on hardware and software trigger efficiencies and on the geometrical light collection efficiency.
To study the detector response and assess the evolution of a multitude of operational parameters over the complete data-taking period, an external $\mathrm{^{57}Co}$ source was inserted into the shielded volume every day.
The $\mathrm{122\,keV}$ interactions, together with Compton-scattering events observed during calibration with a $^{137}\mathrm{Cs}$ source, are used to determine radial response profiles for all channels for position reconstruction in the horizontal plane (both in S1 and S2).
As a result, the average light detection efficiency at the operating field is $\mathrm{1.8\pm0.1}$ and $\mathrm{1.3\pm0.1\,{phe}/{keV_{ee}}}$ for the FSR and SSR, respectively.
Applying the response profiles for each PMT to Monte Carlo simulations of nuclear recoils, one can determine the fraction of events in each bin of reconstructed energy meeting the required 3-fold coincidence.
Alternatively, relaxing the coincidence requirement to 2-fold and studying the fraction of 2-fold to 3-fold events in each bin and comparing with Monte Carlo simulations gives a similar detection efficiency.
The resulting detection efficiencies are shown in Fig.\,\ref{fig:data_MC_leff}.\\

The daily calibrations are also used to determine further correction factors.
For example, the depth-dependence of the ionisation signal S2 varies with the mean electron lifetime in the liquid, which must be corrected accurately.
The width of the electroluminescence signal is proportional to the thickness of the gas phase and hence its polar distribution depends on the detector tilt, which has been observed to vary slowly due to geological factors.
Additionally, the electroluminescence yield varies with gas pressure.
All of these corrections are taken into account.\\

The scintillation yield of liquid xenon depends on the applied electric field.
The recombination of ionisation is partly suppressed by the electric field, resulting in a reduced scintillation response.
The scintillation quenching for electron recoils, $\mathrm{S_{ee}}$, has been measured in {ZEPLIN--III} \cite{Edwards:2009wu} and found to agree with other published data \cite{Aprile:2005ge, Aprile:2006ho}.
In the FSR configuration, $\mathrm{S_{ee} = 0.38}$; for the slightly lower field configuration during the SSR, $\mathrm{S_{ee} = 0.39}$.
The field-induced quenching factor for nuclear recoils has been measured by other experiments and is found to neither change significantly with field nor be very dependent on the nuclear recoil energy \cite{Aprile:2005ge, Manzur:2010bf}.
The value for $\mathrm{S_{nr}}$ is taken to be $0.92$.\\

According to Eq.\,\ref{qyeq}, the total number of electrons extracted into the gas phase for a given signal S2 can be determined with the help of the detector response to single electrons extracted from the liquid.
The signal size of single electrons, $\mathrm{SE}$, has been measured consistently and with high efficiency from two types of data: random triggers and searches for photoionisation between S1 and S2: $\mathrm{SE} = \mathrm{30.6\pm0.5\,phe/e^-}$ in the FSR and $\mathrm{SE} = \mathrm{11.8\pm0.4\,phe/e^-}$ in the SSR \cite{Santos:2011ui, Edwards:2008eh}. 
The electron emission probability at the liquid/gas interface is $\eta = 0.83$ (FSR) and $0.66$ (SSR)\cite{Gushchin:1979ta}.

\section{Fitting and uncertainties}
\subsection{Relative scintillation efficiency}
\begin{figure}[tb]
\includegraphics[width=\linewidth]{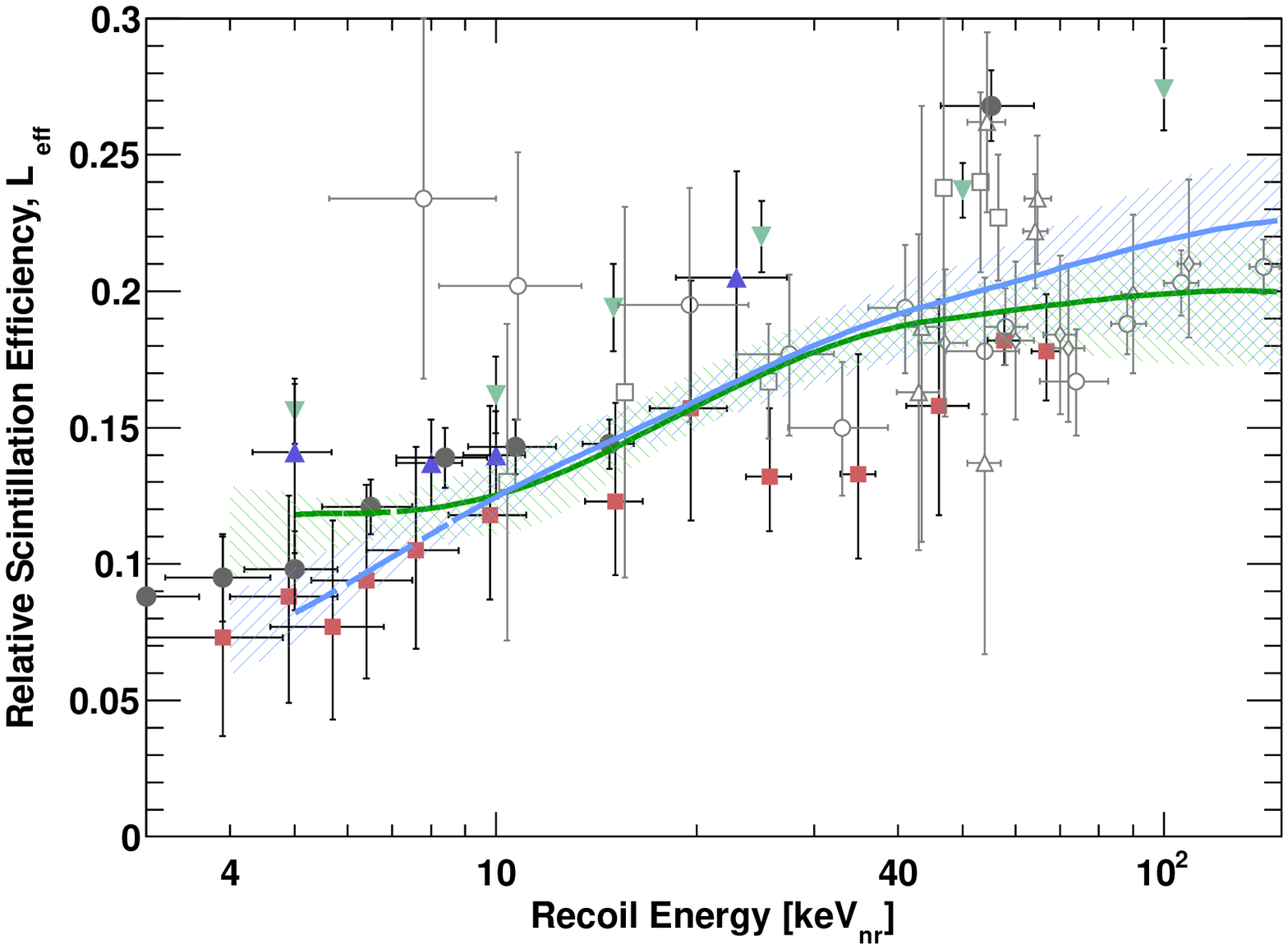}
\caption{\label{fig:l_eff}
The energy-dependent relative scintillation yield for nuclear recoils \leff (solid lines) for the two {ZEPLIN--III} datasets, including relevant 68\% C.L. bands (FSR: green \protect\yout{\,\;}, SSR: blue \protect\xout{\,\;}).
Below the analysis range (corresponding to $\sim\,7-9\,$\protect\kevnr), the scintillation yield is indicated as dashed lines. 
Also shown are previous published measurements using mono-energetic neutron beams:  
($\newmoon$)\cite{Plante:2011hw},
($\blacksquare$)\cite{Manzur:2010bf},
($\filledmedtriangleup$)\cite{Aprile:2009fi}, 
($\fullmoon$)\cite{Arneodo:2000co}, 
($\medtriangleup$)\cite{Akimov:2002vp},
($\medtriangledown$)\cite{Chepel:2006ik} and 
($\meddiamond$)\cite{Aprile:2006ho},
or obtained using a similar Monte Carlo matching procedure ($\filledmedtriangledown$)\cite{Sorensen:2009ec}.}
\end{figure}
The nuclear recoil energy spectrum obtained from the Monte Carlo simulations (in \kevnr) can be converted to a spectrum in reconstructed energy $\mathrm{\langle S1\rangle}$ (in \kevee) via Eq.\,\ref{leffeq}.
At low energy, the energy resolution applied to the Monte Carlo is defined by Poisson fluctuations in the number of photoelectrons and PMT gain fluctuations and was modelled by a continuous Poisson distribution replacing the factorial by Euler's $\Gamma$ function.
The resulting convolved energy spectrum is multiplied by the overall detection efficiency and compared with the experimental Am-Be data (see Fig.\,\ref{fig:data_MC_leff}).
To allow optimal freedom and avoid any model bias, natural piece-wise cubic splines with continuous 1st and 2nd derivatives are used to parameterise the energy dependence of \leff, a method similar to that used in Ref.\,\cite{Sorensen:2009ec}.
The global best fit to the data in the energy range between $2.0\,$\kevee (FSR), or $2.5\,$\kevee (SSR), and $100\,$\kevee\ is found by using a minimum $\chi^2$ technique and is shown in Fig.\,\ref{fig:l_eff}.
The lower analysis thresholds are determined by an approximately 50\% detection efficiency.
In this work, the spline points are fixed at $0.5, 2.5, 6.0, 15, 50$ and $200\,$\kevnr\ and are unconstrained in \leff.
The outcome of the fit has proven to depend very little on the position and number of spline points (since the best fit functions are almost featureless).
The dashed lines in Fig.\,\ref{fig:data_MC_leff} show the differential rate as a function of the reconstructed energy of the Monte Carlo data using the derived form of \leff \,for the two data sets.

\subsection{Recoil ionisation yield}
\begin{figure}[tb]
$\begin{array}{c}
	\includegraphics[width=\linewidth]{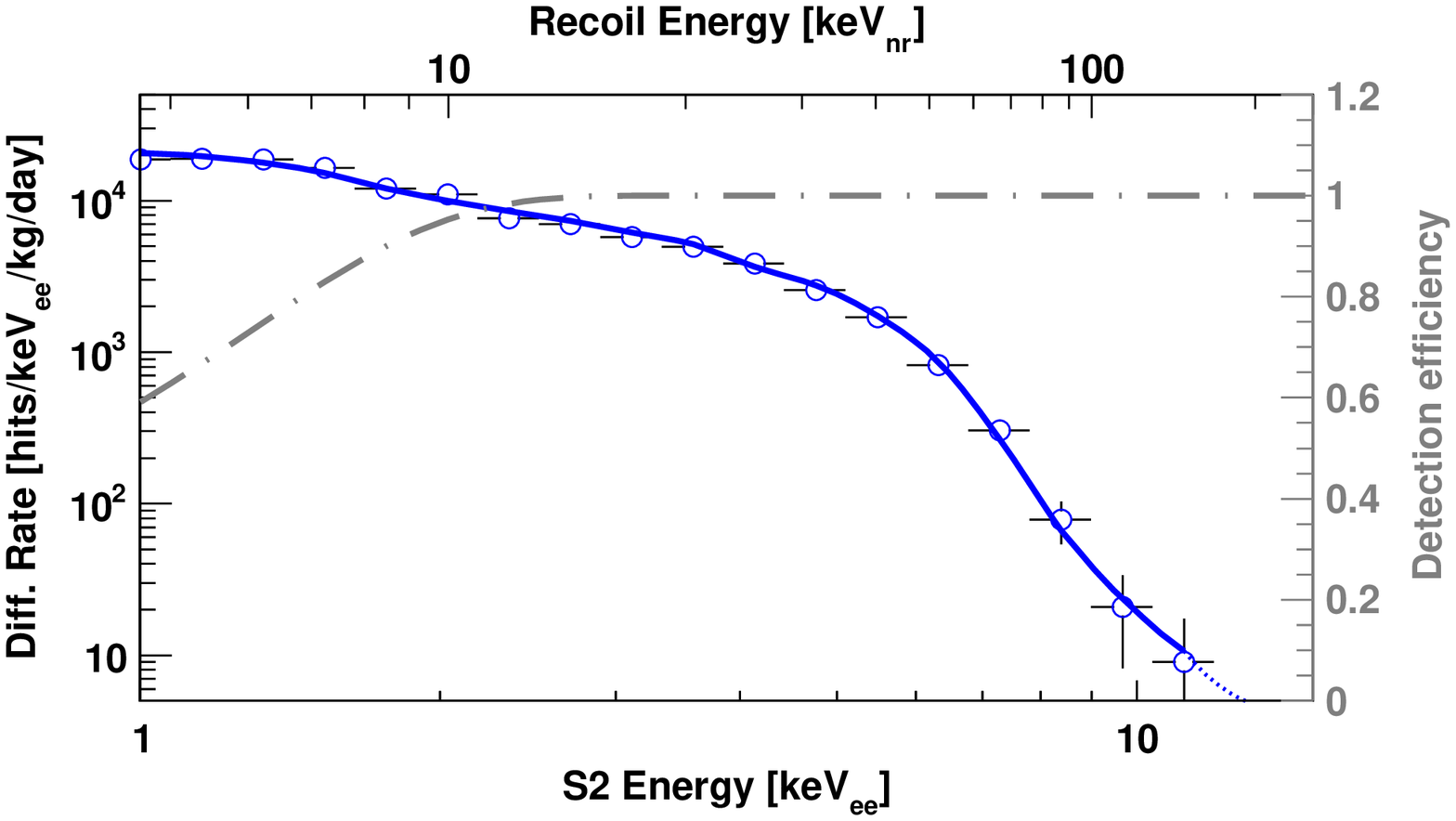} \\
	\includegraphics[width=\linewidth]{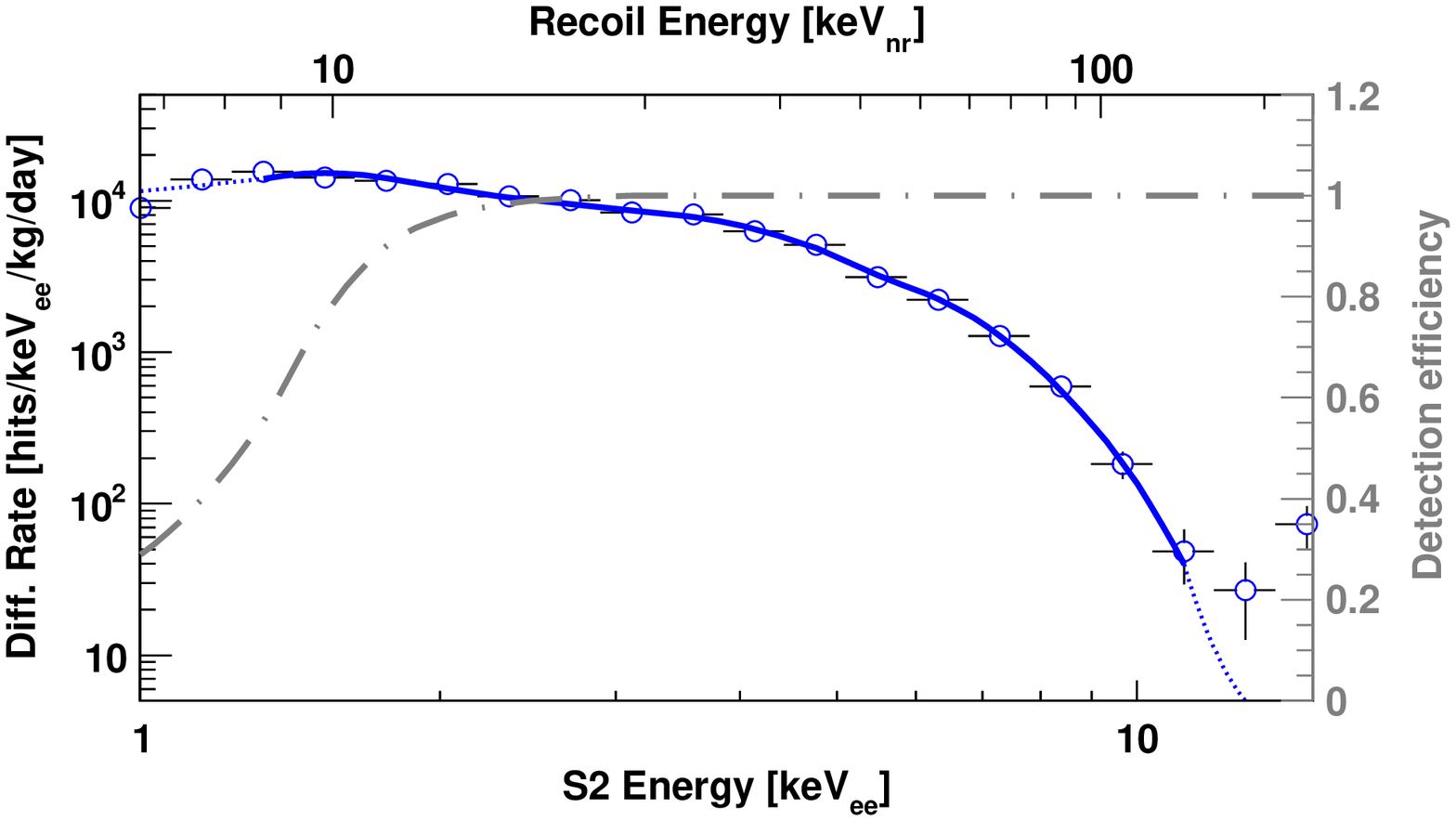}
\end{array}$
\caption{\label{fig:data_MC_qy}
The differential spectrum of secondary scintillation signals from single scatter nuclear recoils from the two datasets (top: FSR; bottom: SSR). 
Also shown are the Monte Carlo data (solid lines) considering an energy dependence of  \qy\ as in Fig.\,\ref{fig:q_y} and the corresponding nuclear recoil energy scale. 
S2 energies below $1.0\,$\kevee (FSR)  or $1.2\,$\kevee (SSR) and above $11\,$\kevee\ are not used to obtain the fits and are shown only as extrapolations (dotted). Superimposed are the detection efficiencies (dash-dotted) as obtained by the primary scintillation signal S1 (see Fig.\,\ref{fig:data_MC_leff})  and converted into S2 energy using the mean correlation between the two channels.}
\end{figure}
\begin{figure}[tb]
\includegraphics[width=\linewidth]{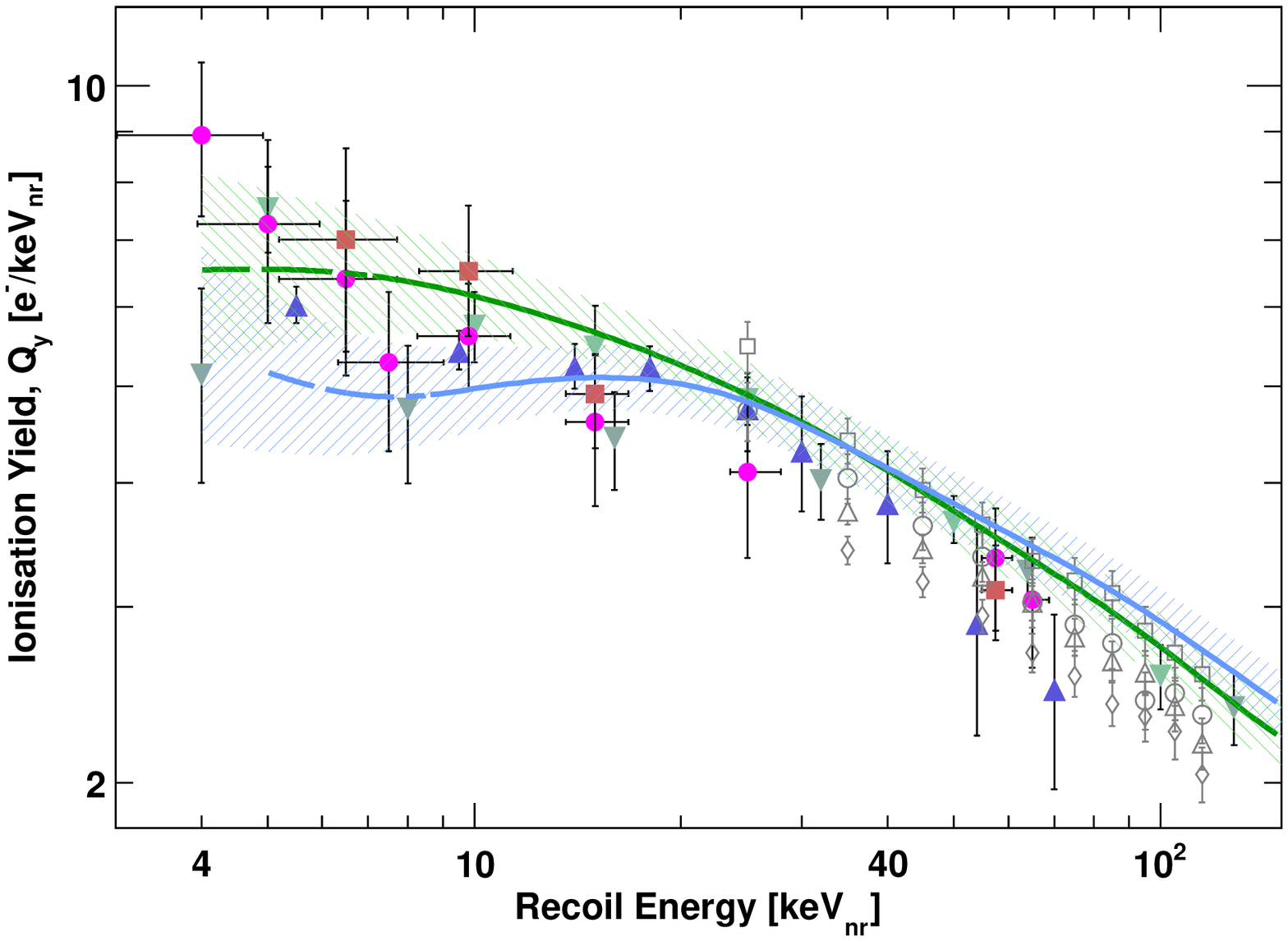}
\caption{\label{fig:q_y} The ionisation yield $\mathcal{Q}_{\mathrm{y}}$ for nuclear recoils as derived from the FSR (green \protect\yout{~\;}) and SSR (blue \protect\xout{~\;}) datasets including relevant 68\% C.L. bands.
Results below the analysis range are indicated by dashed lines.
Also shown are previous measurements 
at $1.0\,\mathrm{kV/cm}$ ($\newmoon$) and $4.0\,\mathrm{kV/cm}$ ($\blacksquare$) from Ref.\,\cite{Manzur:2010bf}, 
at $2\,\mathrm{kV/cm}$ ($\square$), ($\medtriangleup$), 
$0.3\,\mathrm{kV/cm}$ ($\fullmoon$) and $0.1\,\mathrm{kV/cm}$ ($\meddiamond$) from Ref.\,\cite{Aprile:2006ho} 
and spectra obtained using similar Monte Carlo matching procedures at $0.73\,\mathrm{kV/cm}$ ($\filledmedtriangleup$)\,\cite{Sorensen:2009ec} and
($\filledmedtriangledown$)\,\cite{Sorensen:2010vf}.}
\end{figure}
A similar approach to determine \leff, is also applied to the energy dependence of \qy.
In general the W-value is defined as the average energy required to produce an electron-ion pair.
Using the number of electrons recorded in S2 for $122\,$\kevee\ $\gamma$-rays from $\mathrm{^{57}Co}$, we can confirm the W-value for liquid xenon by rearranging Eq.\,\ref{qyeq}. 
For SSR data we obtain $\mathrm{W = 16.5\pm0.8\,eV}$, which agrees very well with the reference measurement of $\mathrm{W = 15.6\pm0.3\,eV}$ published in Ref.\,\cite{Takahashi:1975fz}.
A rate-dependent photocathode charging effect for the FSR PMTs prevented an accurate W-value being derived from the FSR data.
This effect has not been seen in the primary scintillation signal S1, and hence is irrelevant for the determination of the energy scale for low energy recoils and does not influence the results reported in Ref.\,\cite{Lebedenko:2009ce}.
For the following analysis of the ionisation yield and the comparison to the SSR measurement, the W-value was set to $\mathrm{W = 15.6\,eV}$ for this dataset.\\

The observed nuclear recoil spectra for Am-Be neutrons are converted to number of electrons via Eq. \ref{qyeq} and convolved using Gaussian deviates into a number of electrons emitted from the liquid.
Applying the same technique as for \leff, the obtained spectra are then matched to the data by varying the energy dependence of \qy. 
Spline points are fixed at $0.5, 4.0, 10, 30, 75$ and $250\,$\kevnr.
Fig.\,\ref{fig:data_MC_qy} shows the differential rate in {ZEPLIN--III} and the relevant detection efficiency for the FSR and SSR data.
In contrast to the scintillation signal, the overall detection efficiency for the ionisation signal is predominantly determined by S1, since the analysis of data requires a primary signal.
For high nuclear recoil energies, both S1 and S2 are detected without any efficiency losses.
Below S1 signals of $\sim$5\,\kevee, corresponding to S2 signals for nuclear recoils of $\sim$2\,\kevee, the correlation between ionisation and scintillation signal becomes unreliable due to the low number of S1 photoelectrons.
Hence, to avoid any dependence on the effective scintillation yield, \leff, the ionisation yield is only determined for  nuclear recoil energies above $10\,$\kevnr\ and the extension to lower energies is indicative only.
The result is shown in Fig.\,\ref{fig:q_y}, together with results of a similar approach by {XENON10} \cite{Sorensen:2009ec, Sorensen:2010vf}, as well as measurements at various electric fields and nuclear recoil energies using fixed neutron energy scattering experiments \cite{Aprile:2006ho, Manzur:2010bf}.

\subsection{Alternative method to determine \leff}
\begin{figure}[tb]
\includegraphics[width=\linewidth]{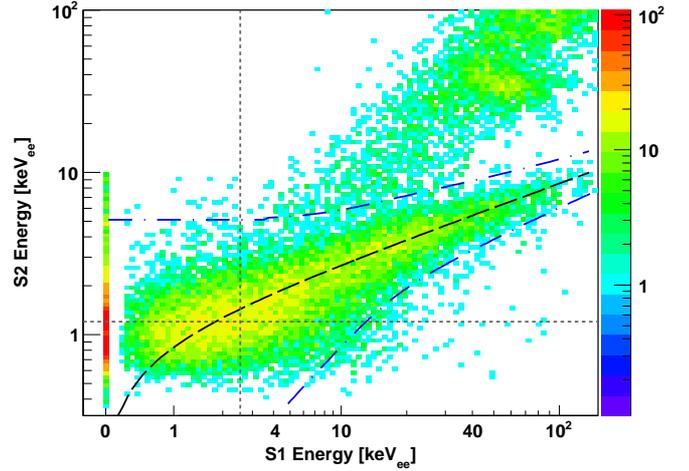}
\caption{\label{fig:belov} The nuclear recoil response to Am-Be neutrons in the SSR configuration including events below the previous analysis threshold for S1 and S2 (dotted lines). The colour scale gives the counts for a 5.9h exposure. Also indicated are the $\pm$\,3\,$\sigma$ contours to the centroid of the neutron recoil population (dashed and dashed-dotted) and their extension sub-threshold.}
\end{figure}
\begin{figure}[tb]
\includegraphics[width=\linewidth]{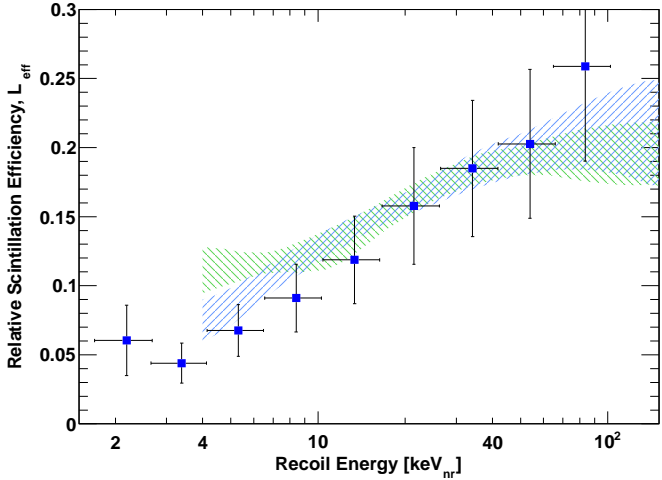}
\caption{\label{fig:belov2} The relative scintillation yield for nuclear recoils $\mathcal{L}_{\mathrm{eff}}$ as determined by extending the event selection to S2-only events for the SSR Am-Be data. Also shown are the results obtained using the Monte Carlo fitting method, as in Fig.\,\ref{fig:l_eff}.}
\end{figure}
In order to provide a cross-check to the scintillation yield, \leff, a further analysis has been developed omitting the dependency on the scintillation detection efficiency on the Monte Carlo simulations and relying instead on ionisation yield only.
This has been achieved by relaxing the event selection criteria to include any events with an ionisation signal and, hence, lowering the energy threshold for nuclear recoils and avoiding detection efficiency penalties in S1.
The scintillation signal for a given nuclear recoil energy in this analysis is estimated from the average amount of light preceding the ionisation signal, regardless of the 3-fold coincidence requirement.
This results in an effective threshold for the scintillation signal close to zero.
Assuming a power law fit to the experimental data of Fig.\,\ref{fig:q_y}, the number of electrons of the ionisation signal can be converted to the corresponding nuclear recoil energy using Eq.\,\ref{qyeq}.
The effective scintillation light yield is then determined via Eq.\,\ref{leffeq}.
In Fig.\,\ref{fig:belov} the event selection for nuclear recoils in this additional analysis is shown, including sub-threshold events.
Here, only events within $\mathrm{120\,mm}$ from the centre of the detector are considered to avoid any bias due to lower light collection of peripheral events.
Due to the lack of the primary scintillation signal and, hence, resolution in the z-axis, the correction due to the mean electron lifetime in the liquid is a depth-averaged value for low energy nuclear recoils.
This is justified by the reasonably homogenous distribution of these recoil events in the xenon volume.\\

To avoid bias from PMT thermionic photoelectron emission, cross-phase single electron emission and other spurious effects producing non-correlated {VUV} light, an average background is determined by integrating all pulses over the same time period {\em before} primary scintillation signals (S1).
This background was found to be $3.1\pm0.1\,\mathrm{phe/event}$ during the Am-Be calibration in the SSR.
The results of this alternative analysis method are shown in Fig.\,\ref{fig:belov2}.
The errors arise predominantly from the Poisson fluctuations in number of photoelectrons and ionisation electrons, as well as from the light collection efficiency and include the effect due to the fitting of a power law for \qy.
To summarise, this method was used to derive the effective scintillation yield, \leff, without relying on the Monte Carlo simulations or being limited by detection efficiencies and the results are in agreement with the spectrum fitting analysis method.

\subsection{Error analyses}
The $\mathrm{68\%}$ confidence intervals, as shown in Fig.\,\ref{fig:l_eff} and \ref{fig:q_y}, are determined by models of \leff\ and \qy\ giving $\chi^2 < \chi^2_{min} + Q_{\gamma}$, where $Q_{\gamma} = 7.01$ for six free parameters \cite{cowan:1998ut}.
Another uncertainty beyond the fitting procedure contributing to the results of the scintillation and ionisation yields arise from the detector dead-time and the Am-Be neutron source strengths.
Both of these parameters scale directly the simulated recoil spectra.
The neutron rates from the two sources are known within 6.5\% (FSR) and 1.1\% (SSR) and the dead-time of the system produces an uncertainty of $\sim$\,2$\%$ in the live times of the neutron exposures for the two runs.
These uncertainties propagate into $<$\,5$\%$ change of \leff\ and \qy\ over the analysed energy range.\\

The results are also affected by systematics of the detector and our knowledge of the properties of liquid xenon.
For example, the photoelectron yield of the chamber, $\mathrm{L_y}$, varies with the position of the interaction in the detector and is determined by $\mathrm{^{57}Co}$ calibration for $\mathrm{122\,keV}$ $\gamma$-rays.
The light collection decreases by $\sim$\,10$\%$ of the volume-averaged value for interactions more than $\mathrm{100\,mm}$ away from the centre of the detector.
However, including more peripheral interactions in this analysis (resulting in a lower average yield, but more events) did not change the results for the effective scintillation yield \leff by more than a few percent.
These are the dominant systematic uncertainties in this analysis.
Their combined magnitude does not contribute beyond the statistical error estimation from the fitting procedure as shown in Fig.\,\ref{fig:l_eff}.
Other systematic errors, also affecting the beam measurements, are discussed critically in Ref.\,\cite{Manalaysay:2010mb}.\\

Similar error estimations have been done for the ionisation yield measurements beyond the statistical $68\%$ C.L. of the fitting procedure.
Here, the detection efficiency as shown in Fig\,\ref{fig:data_MC_qy}, is afflicted by the aforementioned correlation between S1 and S2 below $\sim$\,10\,\kevnr\ and the significant relative variance of photoelectron numbers at these energies.
Further, the emission efficiency of electrons from the liquid into the gas phase is strongly dependent on the electric field in the liquid and thus, in the case of {ZEPLIN--III} on the precise thickness of the liquid volume.
A conservative systematic error of 5\% on $\eta$ directly translates to the ionisation yield \qy\ via the W-value.
In the FSR, the uncertainty on the used measurement of the W-value (2\%) from Ref.\,\cite{Takahashi:1975fz} has been added to the systematics.
For \qy, the combined systematic errors exceed the statistical errors from the fitting procedure and hence are included in the error band in Fig.\,\ref{fig:q_y}.

\section{Conclusions}
We have presented the energy dependence of the scintillation and ionisation yields of nuclear recoils above $10\,$\kevnr\ as seen in the {ZEPLIN--III} experiment using two data sets.
The neutron calibration data obtained during the first science run (FSR) in 2008 were reanalysed using the same pulse finding and position reconstruction algorithms as used on the 2010 data.
The reanalysis of the FSR data revealed a very small, but critical population of events whose analysis relied on the dual-range data acquisition cross-over point.
This explains the new \leff\ curve.
The WIMP-search results are only marginally affected with now 5 events in the (wider) 7--36\,\kevnr\ signal acceptance region. 
This corresponds to an electron recoil leakage of $1:7800$ in the worst case. 
The 90\% C.L. limit on the WIMP-nucleon scalar cross-section improves for WIMP masses below $30\,\mathrm{GeV/c^2}$ and increases for higher masses relative to that published in Ref.\,\cite{Lebedenko:2009ce}.
For example at the bottom of the sensitivity curve ($\sim$\,50\,$\mathrm{GeV/c^2}$) the limit is 23\% higher.\\

Using a complete new set of PMTs in the second science run configuration (with a 40-fold reduction in radiological background and very different optical and electrical performance), a slightly lower external electric field and a different neutron source strength and location, consistent results were obtained for the scintillation yield.
Systematic errors resulting from the detector configuration and the applied fitting method are understood and show only an overall shift within a few percent, and result in little change of the spectrum.
The results below $10\,$\kevnr\ are affected by the overall detection efficiency, the average light yield and light collection efficiency of the {ZEPLIN--III} detector and should be treated as extrapolations only.

\section{Acknowledgements}
\input acknowledgement.tex

\bibliographystyle{model1a-num-names}
\bibliography{leff_biblio}

\end{document}

%% file: author_elsa.tex
\author[ic]{M.~Horn\corref{cor1}}
\ead{m.horn@imperial.ac.uk}

\author[itep]{V.~A.~Belov}
\author[itep]{D.~Yu.~Akimov}
\author[ic]{H.~M.~Ara\'ujo}
\author[edi]{E.~J.~Barnes}
\author[itep]{A.~A.~Burenkov}
\author[lip]{V.~Chepel}
\author[ic]{A.~Currie}
\author[ral]{B.~Edwards}
\author[edi]{C.~Ghag}
\author[edi]{A.~Hollingsworth}
\author[ral]{G.~E.~Kalmus}
\author[itep]{A.~S.~Kobyakin}
\author[itep]{A.~G.~Kovalenko}
\author[ic]{V.~N.~Lebedenko\fnref{fn1}}
\author[lip, ral]{A.~Lindote}
\author[lip]{M.~I.~Lopes}
\author[ral]{R.~L\"{u}scher}
\author[ral]{P.~Majewski}
\author[edi]{A.~StJ.~Murphy}
\author[lip,ic]{F.~Neves}
\author[ral]{S.~M.~Paling}
\author[lip]{J.~Pinto da Cunha}
\author[ral]{R.~Preece}
\author[ic]{J.~J.~Quenby}
\author[edi]{L.~Reichhart}
\author[edi]{P.~R.~Scovell}
\author[lip]{C.~Silva}
\author[lip]{V.~N.~Solovov}
\author[ral]{N.~J.~T.~Smith}
\author[ral]{P.~F.~Smith}
\author[itep]{V.~N.~Stekhanov}
\author[ic]{T.~J.~Sumner}
\author[ic]{C.~Thorne}
\author[lip]{L.~de Viveiros}
\author[ic]{R.~J.~Walker}

\address[ic]{High Energy Physics group, Blackett Laboratory, Imperial College London, UK}
\address[itep]{Institute for Theoretical and Experimental Physics, Moscow, Russia}
\address[edi]{School of Physics \& Astronomy, SUPA University of Edinburgh, UK}
\address[lip]{LIP--Coimbra \& Department of Physics of the University of Coimbra, Portugal}
\address[ral]{Particle Physics Department, STFC Rutherford Appleton Laboratory, Chilton, UK}

\cortext[cor1]{Corresponding author}
\fntext[fn1]{Deceased}

%% file: acknowledgement.tex
%
The UK groups acknowledge the support of the Science \& Technology Facilities Council (STFC) for the ZEPLIN--III project and for maintenance and operation of the underground Palmer laboratory which is hosted by Cleveland Potash Ltd (CPL) at Boulby Mine, near Whitby on the North-East coast of England.  
The project would not be possible without the co-operation of the management and staff of CPL. 
We also acknowledge support from a Joint International Project award, held at ITEP and Imperial College, from the Russian Foundation of Basic Research (08-02-91851 KO a) and the Royal Society.
LIP--Coimbra acknowledges financial support from Funda\c c\~ao para a Ci\^encia e Tecnologia (FCT) through the project-grants CERN/FP/109320/2009 and CERN/FP/116374/2010, as well as the postdoctoral grants SFRH/BPD/27054/2006, SFRH/BPD/47320/2008 and SFRH/BPD/63096/2009.
This work was supported in part by SC Rosatom, contract $\#$H.4e.45.90.11.1059 from 10.03.2011.
The University of Edinburgh is a charitable body, registered in Scotland, with the registration number SC005336.
%